\RequirePackage[2020-02-02]{latexrelease}
\documentclass[fleqn,pra,floatfix,showkeys,preprint,nofootinbib]{revtex4}

\usepackage{graphicx}
\usepackage{amsmath}
\RequirePackage[2020-02-02]{latexrelease}
\usepackage{dcolumn}
\usepackage[cp1252]{inputenc}
\begin{document}

\title{Experimental determination of the energy dependence
of the rate of the muon transfer reaction from muonic hydrogen to oxygen 
for collision energies up to 0.1 eV}

\author{M.~Stoilov$^3$}
\email{mstoilov@inrne.bas.bg}

\author{A.~Adamczak$^2$}
\author{D.~Bakalov$^3$}%
\author{P.~Danev$^3$}%
\author{E.~Mocchiutti$^1$}%
\author{C.~Pizzolotto$^1$}
\author{G.~Baldazzi$^4$}
\author{M.~Baruzzo$^{5,1}$}
\author{R.~Benocci$^{6,7}$}
\author{M.~Bonesini$^{6,7}$}
\author{D.~Cirrincione$^{1,5}$}
\author{M.~Clemenza$^{6,7}$}
\author{F.~Fuschino$^{4,11}$}
\author{A.D.~Hillier$^{12}$}
\author{K.~Ishida$^{13}$}
\author{P.J.C.~King$^{12}$}
\author{A.~Menegolli$^{9,10}$}
\author{S.~Monzani$^{5,1}$}
\author{R.~Ramponi$^{14,15}$}
\author{L.P.~Rignanese$^4$}
\author{R.~Sarkar$^{16}$}
\author{A.~Sbrizzi$^{4}$}
\author{L.~Tortora$^8$}
\author{E.~Vallazza$^6$}
\author{A.~Vacchi$^{1,5,13}$}

 \affiliation{$^1$ Sezione INFN di Trieste, via A. Valerio 2,
 Trieste, Italy}

 \affiliation{$^2$ Institute of Nuclear Physics, Polish Academy of
  Sciences, Radzikowskiego 152, PL31342 Krak\'{o}w, Poland}

 \affiliation{%
 $^3$ Institute for Nuclear Research and Nuclear Energy, Bulgarian Academy of Sciences,\\
    blvd. Tsarigradsko ch. 72, Sofia 1142, Bulgaria}

 \affiliation{$^{4}$ Sezione INFN di Bologna, viale Berti Pichat 6/2, Bologna, Italy}

 \affiliation{$^{5}$ Dipartimento di Scienze Matematiche, Informatiche e Fisiche, Universit\`{a} di Udine, via delle Scienze 206, Udine, Italy}

 \affiliation{$^{6}$ Sezione INFN di Milano Bicocca, Piazza della Scienza 3, Milan, Italy}

 \affiliation{$^{7}$ Dipartimento di Fisica G. Occhialini, Universit\`{a} di Milano Bicocca, Piazza della Scienza 3, Milan, Italy}

 \affiliation{$^{8}$ Sezione INFN di Roma Tre, Via della Vasca Navale 84, Rome, Italy}

 \affiliation{$^{9}$ Dipartimento di Fisica, Universit\`{a} di Pavia, via A. Bassi 6, Pavia, Italy}

 \affiliation{$^{10}$ Sezione INFN di Pavia, Via A. Bassi 6, Pavia, Italy}

 \affiliation{$^{11}$ INAF-OAS Bologna, via P. Gobetti 93/3, Bologna, Italy}

 \affiliation{$^{12}$ ISIS Neutron and Muon Source, STFC Rutherford-Appleton Laboratory, Didcot OX11 0QX, UK}

 \affiliation{$^{13}$ Riken Nishina Center, RIKEN, 2-1 Hirosawa, Wako, Saitama 351-0198, Japan}

 \affiliation{$^{14}$ Sezione INFN di Milano, via Celoria 16, Milan, Italy}

 \affiliation{$^{15}$ IFN-CNR, Dipartimento di Fisica, Politecnico di Milano, piazza Leonardo da Vinci 32, Milan, Italy}

 \affiliation{$^{16}$ Indian Centre for Space Physics, Kolkata, India,}

 \date{\today}

 \begin{abstract}
We report the first experimental determination of the collision-energy dependence of the muon transfer rate from the ground state of muonic hydrogen to oxygen at near-thermal energies. A sharp increase by nearly an order of magnitude in the energy range 0 - 70 meV was found that is not observed in other gases. The results set a reliable reference for quantum-mechanical calculations of low-energy processes with exotic atoms, and provide firm ground for the measurement of the hyperfine splitting in muonic hydrogen and the determination of the Zemach radius of the proton by the FAMU collaboration.
 
 \end{abstract}


\maketitle


 \section{Introduction}
 \label{sec:intro}

  Muon transfer in collision of muonic hydrogen with a heavier atom is an example of charge transfer in non-elastic scattering of ion $A^+$ by atom $B$:
  \begin{align}
  A^+ + B \rightarrow A + B^+.
  \label{eCT}
  \end{align}
  Charge transfer reactions with exchange of an electron are a broad class of processes that have been extensively studied for decades both theoretically and experimentally. A general outlook on the topic could be found, e.g. in \cite{may} and the references therein; for recent advances see  \cite{review1}. Here we only mention the investigations of charge transfer in scattering of those light atoms and ions, 
  the ``muonic'' counterparts of which have been studied experimentally (see next paragraph).
 The energy dependence of the charge transfer rate in argon-nitrogen scattering at near-thermal collision energies was studied in \cite{lindinger,rebrion,candori}; Refs.~\cite{lindsay,lindsay1} were focused on charge transfer at epithermal energies.  Ref.~\cite{dalgarno} presents a thorough theoretical study of the energy dependence of charge transfer between hydrogen isotopes at low energies and the validity of Wigner law;
the latter is discussed in full details in Ref.~\cite{sadeg}.

Muonic atoms are formed when negative muons are stopped in matter and captured by the Coulomb field of the nuclei -- initially in an excited state, which is promptly de-excited 
via a set of competing mechanisms including Auger effect (for higher Z muonic atoms), Stark mixing, collisional Coulomb de-excitation etc. \cite{wu,ponom,markushin}. The de-excitation steps are signaled by the emission of characteristic X-rays \cite{lauss}. 
Muonic hydrogen is a special case: the muon replaces the only atomic electron, 
and because of the large muon mass $(m_{\mu}/m_e\sim207)$ 
and its small size (characteristic length scale $a_{\mu}\sim a_{\infty}(m_e/m_{\mu})\sim 0.26\times10^{-10}$ cm),
the muonic hydrogen atom in the ground $1s$ state behaves, at the Bohr radius 
$a_{\infty}$ scale, as a neutral particle. This allows the atom $\mu^-p_{1s}$ to penetrate the electronic cloud of higher-Z atom $X$ and transfer the muon to the nucleus in an analog of the electron exchange reaction (\ref{eCT}):
  \begin{align}
  \mu^-p + X^{Z+} \rightarrow p + (\mu^-X)^{(Z-1)+}.
  \label{muCT}
  \end{align}
Though similar, the reactions of muon (\ref{muCT}) and electron (\ref{eCT}) transfer differ in many aspects. Muon transfer from muonic hydrogen is essentially a three-body process, and the influence of the electron structure of the higher-Z atom consists  mainly in screening the Coulomb field of its nucleus.

This has necessitated the development of new methods for the quantitative theoretical description of process (\ref{muCT}).
Calculations of the rate of muon transfer from muonic hydrogen to light higher-Z atoms at thermal and epithermal energies have been carried out
with increasing accuracy using classical trajectories \cite{haff}, in the
adiabatic approach \cite{adiabat,adiab1,adiab2}, semiclassically \cite{gersht63,fior,belyaev}, in the WKB approximation \cite{WKB}, using Faddeev-Hahn equations \cite{sultanov}, within the method of perturbed stationary states \cite{romanov,romanov22}, and in the hyperspherical approach \cite{dupays1,dupays2,dupays3,tcherbul,cdlin,igarashi}. On the experimental side, the
scarce amount of muonic hydrogen atoms (as compared with charge transfer experiments with electronic atoms), and the short lifetime of the muon, required the use of  techniques inspired by experimental particle physics, such as the analysis of the time evolution of the characteristic X-ray spectra. The measurements, performed at fixed, predominantly room temperature, in a mixture of hydrogen and higher-Z gases, have provided the  rate of muon transfer from hydrogen to helium \cite{bystr,gartner,tresch},  carbon \cite{piller}, nitrogen, neon, and argon \cite{thalmann1,thalmann2,jacot-neon,jacot-neon1}, and oxygen \cite{werth0,werth} at thermal energy. 
Estimates of the muon transfer rate at higher energies were obtained from data on the ``epithermal muon transfer events'' occurring from not-yet-thermalized muonic hydrogen atoms. The observed variations of the rates with energy, pressure and admixture concentrations were  qualitatively explained in the existing models of formation and diffusion of muonic hydrogen atoms, except for the unexpectedly strong dependence on the collision energy of the rate of muon transfer to oxygen \cite{werth0,werth}. The reaction
  \begin{equation}
  p\mu^-+O_2\rightarrow p+(O\mu^-)+O
  \label{eq:transfer}
  \end{equation}
has been attracting the attention of both experimentalists and theorists since the discovery of the double-exponential time spectra of muonic oxygen X-rays \cite{mulhauser}.
The experimental investigations of (\ref{eq:transfer}) in the 90's led to the two-step model \cite{werth0} for the rate of the  process, 
which was consistent with the then available data from measurements at room temperature. The interest in the subject was revived a few years later in  relation to the projects to measure the hyperfine splitting in the ground state of muonic hydrogen \cite{jinst18,epja,japs-las,crema-las,crema-new},
  and extract out of it the value of the electromagnetic
  Zemach radius of the proton \cite{ours,cjp}.
  In a series of advanced
  theoretical calculations \cite{dupays1,cdlin,romanov22} significant progress
  was achieved in the quantitative description of the process
  (\ref{eq:transfer}) for energies up to 10 eV, but
  these theoretical results necessitate
  experimental verification.
  The breakthrough came with the recent results of the FAMU
  collaboration \cite{pla20,pla21}, which performed the first
  experimental investigation of the temperature dependence of the rate
  of muon transfer from hydrogen to oxygen.
  The rate of the process 
  was measured with high accuracy at a set of
  temperatures in the range between 70 $K$ and
  336 $K$, and the anticipated dependence
  on the target temperature
  was rigorously confirmed. 
  
  The objective of
  the present work is to extract from these experimental data
  reliable estimates for the dependence of the rate of the muon transfer process
  (\ref{eq:transfer}) on the collision energy $E$. The motivation
  of our work is two-fold:

1. Reliable experimental data on the energy dependence of the
  rate of (\ref{eq:transfer}) will provide a reference point for
  the computational methods for the accurate quantitative
  description of low-energy scattering of atoms, and  in particular
  -- of charge transfer in atomic collisions.
  While the results in Refs.~\cite{dupays1,cdlin,romanov22} are in qualitative
  agreement with each other, the remaining significant quantitative discrepancy only reaffirms  the need of such reliable references.

2. The experimental method for the measurement of the
  hyperfine splitting in the ground state of muonic hydrogen of the FAMU collaboration \cite{jinst18,epja} exploits
  substantially the anticipated strong energy dependence of the rate of muon transfer from hydrogen to oxygen.
  Modelling the experiment
  requires detailed and verified quantitative information on this
  dependence in the thermal and near epithermal energy range.
  
In what follows the energy dependence of the rate of muon transfer to oxygen will be determined using constrained fits to the FAMU dataset. 
In Sect.~\ref{sect2} we formulate a set of model-independent constraints on the latter, probe a variety of trial functions (TFs) that satisfy these constraints, 
and select a short list of fits on the ground of statistical criteria.  
In Sect.~\ref{sect3} we analyze the uncertainties of the best fit and compare it to the existing theoretical and experimental results. 
In the conclusive Sect.~\ref{sect4} we outline the fields of possible application of the results, in particular - in the experimental determination of the Zemach radius of the proton.

 \section{Determining the energy dependence of the muon transfer rate to oxygen}
 \label{sect2}
 
 \subsection{Atomic vs. molecular scattering of muonic hydrogen}
 \label{sect2A}

In nonelastic scattering of $p\mu$ atoms by oxygen atoms,
the probability $dP$ that the $p\mu$ atom transfers its muon to the oxygen nucleus  
\begin{equation}
  p\mu^-+O^{8+}\rightarrow p+(\mu^-O)^{7+}
  \label{eq:attransfer}
\end{equation}
within the time interval $dt$ may be put in the form
$dP=\lambda_{\rm pO}^{\rm A}\,\phi^{\rm A}\,dt$,
where $\phi^{\rm A}=\rho^{\rm A}/\rho^{\rm LHD}$, $\rho^{\rm A}$ is the 
number density of the oxygen atoms, and $\rho^{\rm LHD}=4.25\times10^{22}\ {\rm cm}^{-3}$ is the number density of the hydrogen atoms in liquid hydrogen (LHD).
The coefficient $\lambda_{\rm pO}^{\rm A}$ is referred to as ``{\em rate (of the reaction) of muon transfer to oxygen nucleus, normalized to LHD}'', 
the normalization being selected to help compare the rates of different processes in a specific-condition-independent way.
The rate 
$\lambda_{\rm pO}^{\rm A}$ is related to the muon transfer reaction cross section 
$\sigma_{\rm pO}^{\rm A}$ by means of 
$\lambda_{\rm pO}^{\rm A}=\sigma_{\rm pO}^{\rm A}\,\rho^{\rm LHD}\,v$, where $v=\sqrt{2E/m}$ denotes the relative velocity of the colliding $p\mu$ atom and oxygen nucleus, $m$ is their reduced mass, and $E$ stands for the the collision energy in the center-of-mass (CM) reference frame.

The mechanism of muon transfer to an oxygen nucleus in nonelastic scattering of $p\mu$ atoms by oxygen molecules (\ref{eq:transfer}) is 
assumed to be the same as in (\ref{eq:attransfer}) since 
the reaction of muon transfer (\ref{eq:transfer}) is essentially a three-body process, which takes place at interparticle distances of the order of $a_{\mu}$ and is only remotely affected  by the molecular structure. 
The probability $dP$ that, in nonelastic scattering by an oxygen molecule, the $p\mu$ atom transfers the muon to a O$_2$ nucleus, has a similar form: 
$dP=\lambda_{\rm pO}\,\phi^{\rm A}\,dt$, where $\lambda_{\rm pO}$ is the 
{\em rate of muon transfer in nonelastic scattering of muonic hydrogen by oxygen molecules, normalized to LHD oxygen density}. 
It is important, however, to clearly distinguish the rates $\lambda_{\rm pO}$ and 
$\lambda_{\rm pO}^{\rm A}$: the experimentally measurable quantity is 
$\lambda_{\rm pO}$, while $\lambda_{\rm pO}^{\rm A}$ can in principle be calculated (apart from computational difficulties) with high accuracy, but not directly measured. Their numerical values are expected to be close but not equal. Coming back to the motivation of the present work (see Sect.~\ref{sec:intro}), we note that the knowledge of $\lambda_{\rm pO}$ is what is needed to verify the FAMU experimental method. The rate $\lambda_{\rm pO}$ can also serve as reference for the computational methods in low-energy scattering theory provided that these methods are extended to account for the effects of molecular structure.

 \subsection{Temperature vs. energy dependence}

In general, the muon transfer rate depends on $E$; we denote the energy-dependent rate by $\lambda_{\rm pO}(E)$. 
The FAMU measurements of the rate 
of muon transfer in scattering of $p\mu$ by oxygen molecules were performed {\em in a fully thermalized gas target}. The rates were measured at $n_d=10$ different temperatures $T_k,k\le n_d$ in the range $70\le T_k\le 336$ K.
In the conditions of thermal equilibrium, the observable rate of muon transfer at temperature $T$,
 $\Lambda_{\rm pO}(T)$,  is related to $\lambda_{\rm pO}(E)$ by means of
 \begin{equation}
 \Lambda_{\rm pO}(T)=\int\limits_0^{\infty} dE\,f_{\rm MB}(E;T)\,\lambda_{\rm pO}(E),
 \label{eq:bas}
 \end{equation}
 where $f_{\rm MB}(E;T)=(2/\sqrt{\pi})(k_BT)^{-3/2}
 \sqrt{E}\,\exp(-E/k_BT)$ is the Maxwell-Boltzmann distribution; $k_B$ is the Boltzmann constant.
Refs.~\cite{pla20,pla21} describe in detail the experimental set-up. 
The experimental values $\Lambda_k=\Lambda_{\rm pO}(T_k),k=1,...,n_d$ that have been  reported there, are summarized in Table~\ref{tab:dataset}.
 \begin{table}[h!]
 \caption{Compliation of the FAMU experimental data, reported in Refs.~
\cite{pla20,pla21}. Experimental rates
 $\Lambda_k=\Lambda_{\rm pO}(T_k),k\le n_d$
 of muon transfer from hydrogen to oxygen at $n_d=10$
 preselected temperatures.
 The values are normalized to liquid hydrogen density (LHD) 
 $4.25\times10^{22}\text{\ cm}^{-3}$.
 $\sigma^{(1,2)}_k$ and $\sigma_k$ denote the statistical,
 systematic, and overall standard errors of the experimental value $\Lambda_k$.}
 \label{tab:dataset}
 \begin{tabular}{c@{\hspace{4mm}}c@{\hspace{4mm}}c@{\hspace{4mm}}c@{\hspace{4mm}}c@{\hspace{4mm}}c@{\hspace{4mm}}c}
 $k$ & $T_k$ & $\Lambda_k$ &
 $\sigma^{(1)}_k$ & $\sigma^{(2)}_k$ & $\sigma_k$ & Source \\
     & $ [K]$ & $ [10^{10}\ s^{-1}]$  & $ [10^{10}\ s^{-1}]$ & $ [10^{10}\ s^{-1}]$ 
     & $ [10^{10}\ s^{-1}]$
     \\ \hline
  \vrule depth 0pt height 12pt width 0pt    
 1 &  70 & 2.67 & 0.40 & 0.32 & 0.51 & Ref.~\cite{pla21}  \\
 2 &  80 & 2.96 & 0.11 & 0.36 & 0.38 & Ref.~\cite{pla21}  \\
 3 & 104 & 3.07 & 0.29 & 0.07 & 0.30 & Ref.~\cite{pla20}  \\
 4 & 153 & 5.20 & 0.33 & 0.10 & 0.34 & Ref.~\cite{pla20}  \\
 5 & 201 & 6.48 & 0.32 & 0.13 & 0.35 & Ref.~\cite{pla20}  \\
 6 & 240 & 8.03 & 0.35 & 0.16 & 0.38 & Ref.~\cite{pla20}  \\
 7 & 272 & 8.18 & 0.37 & 0.17 & 0.41 & Ref.~\cite{pla20}  \\
 8 & 300 & 8.79 & 0.39 & 0.18 & 0.43 & Ref.~\cite{pla20}  \\
 9 & 323 & 8.88 & 0.62 & 0.66 & 0.91 & Ref.~\cite{pla21}  \\
10 & 336 & 9.37 & 0.57 & 0.70 & 1.07 & Ref.~\cite{pla21}
 \end{tabular}
 \end{table}

\begin{figure}
 \begin{center}
 \includegraphics[width=0.9\textwidth]{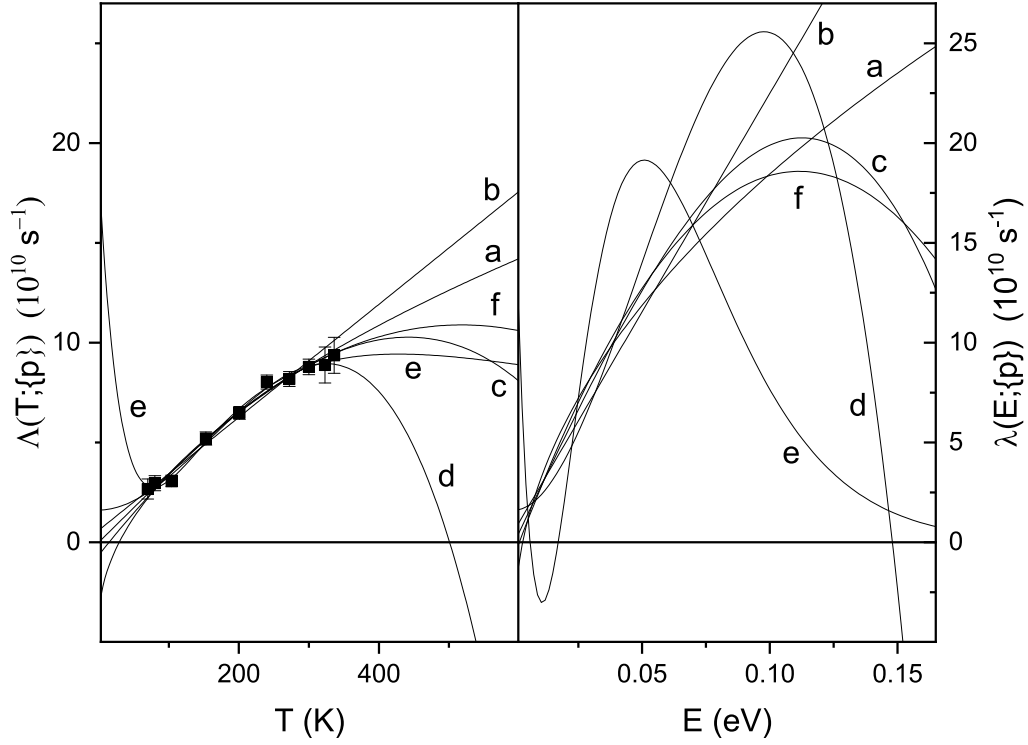}
 \caption{Temperature (left) and energy (right) dependence of the muon transfer rate for a few simple unconstrained fits: (a) $a_0+a_1\sqrt{E},\ \chi^2=7.23$; (b) $a_0+a_1E,\ ,\chi^2=9.39$; (c) $a_0E+a_1E^3,\ \chi^2=5.66$; (d) $a_0+a_1E^2+a_2E^3,\ \chi^2=3.83$; (e) $(a_0+a_1E+a_2E^2)\exp(-E/a_3),\ \chi^2=2.41$; (f) $a_0+a_1E+a_2E^2,\ \chi^2=5.92.$
 The experimental data $\Lambda_k$ are presented with black squares.
The  fits $\Lambda(T;\{p\})$ reproduce reasonably well the data, but diverge outside the range of investigated temperatures, and lead to strongly incoherent energy dependence expressions $\lambda(E;\{p\})$.}
 \label{fig:underdet}
 \end{center}
 \end{figure}

The convolution integral in Eq.~(\ref{eq:bas}) for $\Lambda_{\rm pO}(T)$ may be put in the form of a Laplace transform of $\lambda_{\rm pO}(E)$. If $\Lambda_{\rm pO}(T)$ were known for any $T$ one might obtain $\lambda_{\rm pO}(E)$ by the inverse Laplace transform of $\Lambda_{\rm pO}(T)$. 
The naive approach would be to find a parametric fit\footnote[1]{By $\Lambda(T;\{p\})$ and $\lambda(E;\{p\})$ we denote parametric fits to the (unknown) functions 
$\Lambda_{\rm pO}(T)$ and $\lambda_{\rm pO}(E)$, describing the dependence of the muon transfer rate to oxygen on temperature and energy, respectively.}
 $\Lambda(T;\{p\})$ of the experimental values $\Lambda_k$, and compute $\lambda(E;\{p\})$ as the inverse Laplace transform of $\Lambda(T;\{p\})$. 
This leads, however, to unreliable predictions for the energy dependence 
of the muon transfer rate 
as illustrated on 
Fig.~\ref{fig:underdet}: simple fits that approximate the data reasonably well produce strongly divergent $\lambda(E;\{p\})$, which in some cases even take non-physical negative values. 
The reason is that because of the limited experimental data the inverse problem is ill-posed.
Indeed, the contribution from energies $E\gg k_BT_{10}$ to the integral in the right-hand side of Eq.~(\ref{eq:bas}) is exponentially suppressed that leads to 
exponential growth of the uncertainty of $\lambda(E;\{p\})$, when 
 evaluated 
 at $E\gg k_BT_{10}$.
Similarly, the contribution to the integral from the domain 
$0\le E\le E_0\ll k_BT_1$ decreases as $E_0^{3/2}$ that leads to an increase of the uncertainty of 
$\lambda(E;\{p\})$ as  $E^{-3/2}$ for $E\ll k_BT_1$. 
Having this in mind, we shall derive estimates of $\lambda_{\rm pO}(E)$ following two alternative paths: by applying simple regularization methods for the discretized inverse problem (\ref{sec:reg}), and by exploring appropriately selected classes of constrained parametric fits $\lambda(E;\{p\})$ 
 (\ref{sec:fits}). The comparison of the obtained estimates will serve as an indirect test of their reliability.

\subsection{Regularized solutions}
\label{sec:reg}

To resolve Eq.~(\ref{eq:bas}) for $\lambda_{\rm pO}(E)$ with a regularization method we  discretize the inverse problem by using a Gauss quadrature to approximate the integral in the right hand side with a finite sum.
Possible options are the Gauss-Legendre, Gauss-Laguerre and Gauss-Jacobi quadratures
\cite{abram}; we select the quadrature associated with the Jacobi polynomials $J^{\alpha,\beta}$ with $\alpha=0,\beta=1/2$ to account for the square-root singularity at $E=0$:
\begin{align}
&\Lambda_k=\Lambda_{\rm pO}(T_k)=
\int\limits_0^{\infty} dE\,f_{\rm MB}(E;T_k)\,\lambda_{\rm pO}(E)\approx
\int\limits_0^{E_{\rm max}} dE\,f_{\rm MB}(E;T_k)\,\lambda_{\rm pO}(E)
\label{eq:discr}\\
&=\frac{2}{\sqrt{\pi}}\left(\frac{E_{\rm max}}{2k_B T_k}\right)^{3/2}
\int\limits_{-1}^1 dz\sqrt{z+1}\exp\left(-\frac{E_{\rm max}}{2k_B T_k}(z+1)\right)
\lambda_{\rm pO}(E_{\rm max}(z+1)/2)\nonumber\\
&\approx\frac{2}{\sqrt{\pi}}\left(\frac{E_{\rm max}}{2k_B T_k}\right)^{3/2}
\sum\limits_{i=1}^{n_G} w_i 
\exp\left(-\frac{E_{\rm max}}{2k_B T_k}(z_i+1)\right)
\lambda_{\rm pO}(E_{\rm max}(z_i+1)/2).
\nonumber
\end{align}
where $z_i$ and $w_i$ are the nodes and weights of the Gauss quadrature of rank $n_G$
associated with the Jacobi polynomials $J^{0,1/2}$.
The upper limit $E_{\rm max}$ and the rank $n_G$ are selected to secure the needed accuracy of the truncated integral for any $T_k,k=1,...,n_d$; we probed $0.2\le E_{\rm max}\le0.5$ eV and 
$6\le n_G\le20$, and selected $E_{\rm max}=0.3$ eV and $n_G=n_d=10$. 
The values $\lambda_i$ of the energy dependence function 
$\lambda_{\rm pO}(E)$ at energies $E_i=E_{\rm max}(z_i+1)/2$ are calculated from the linear system 
\begin{align}
&\Lambda_k=\sum\limits_{i=1}^{n_G}A_{ki}\lambda_i,\ k=1,\dots,n_d,\ i=1,n_G\text{, where}
\label{eq:linsys}
\\
&\lambda_i=\lambda_{\rm pO}(E_i),\ \ E_i=E_{\rm max}(z_i+1)/2,\ \ 
A_{ki}=\frac{2}{\sqrt{\pi}}\left(\frac{E_{\rm max}}{2k_B T_k}\right)^{3/2}
w_i\, \exp\left(-\frac{E_i}{k_B T_k}\right).
\nonumber
\end{align}
For $n_G=10$ the matrix $A$ is ill-conditioned and the inverse problem is ill-posed (and underdetermined for $n_G>n_d$). We therefore apply regularization to obtain a reliable approximate solution of (\ref{eq:linsys}).

Denote by $(U,V,D)$ the singular value decomposition of $A$: 
\begin{align}
A=UDV^T,\ U^{-1}=U^T,\ V^{-1}=V^T,\ D=\{D_{ik}\}=\!
\begin{cases}
0, i\ne k,i\le n_G,k\le n_d\\
d_i,i=k\le\min(n_G,n_d)
\end{cases}\!\!\!\!\!\!,
d_i\ge d_{i+1}\ge0.
\nonumber
\end{align}
The minimum norm approximate solution of the regularized problem (\ref{eq:linsys}) is given  
\cite{kaipio} by $\lambda_i=\sum_iA^{\dagger}_{ik}\Lambda_k,\ A^{\dagger}=V D^{\dagger} U^T$, where the explicit form of $D^{\dagger}$ depends on the regularization method. Accordingly, the  estimate of the statistical error $\delta^{\rm st}\lambda_i$ of $\lambda_i$ is given by 
$(\delta^{\rm st}\lambda_i)^2=\sum_k(A^{\dagger}_{ik})^2(\sigma_k)^2$.
We probe two simple regularization method.

\begin{figure}
 \begin{center}
 \includegraphics[width=0.95\textwidth]{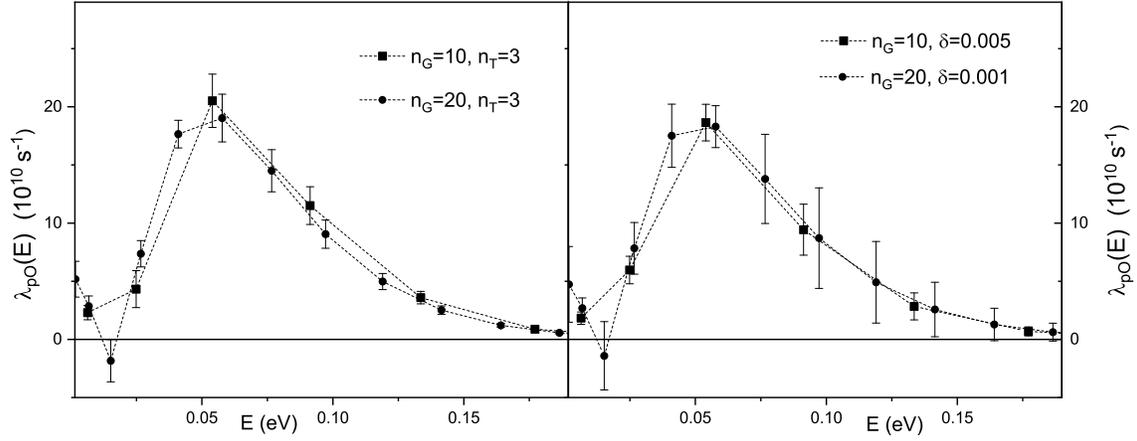}
 \caption{Muon transfer rates $\lambda_i=\lambda_{\rm pO}(E_i)$ calculated as solution of the linear system (\ref{eq:linsys}), regularized by truncating the singular value decomposition at  $n_T=3$ (left) or by Tiknonov's method with appropriate regularization parameter $\delta$ (right). Plotted are the minimum norm solutions for quadrature rank $n_G=10$ and 20 and the estimate of the statistical errors $\delta^{\rm st}\lambda_i$.}
 \label{fig:regfits}
 \end{center}
 \end{figure}

1.  {\em Truncated singular value decomposition regularization} (TSVD). In this case all matrix elements of $D^{\dagger}$ are null except for $D^{\dagger}_{ii}=1/d_i,i\le n_T$.
The truncation level $n_T$ is determined from the discrepancy principle \cite{kaipio};
in our case it turns out $n_T=3$.
On Fig.~\ref{fig:regfits}, left, we plot the values $\lambda_i$ calculated in this way and, for 
comparison, the solution obtained with $n_T=3$ and $n_G=20$. Solutions with $n_G> n_d$ are not positive definite due to the underdeterminedness of (\ref{eq:linsys}). 
Note that the calculations provide only the approximate values of $\lambda_{\rm pO}(E)$ at the energies $E_i,i=1,...,n_G$, of which only a few are within the interval of main interest 
$0\le E\le0.1$ eV; the connecting dashed straight lines serve to distinguish the solutions but do not carry any information about the behavior of $\lambda_{\rm pO}(E)$ between the nodes $E_i$. Increasing the quadrature rank $n_G$ in order to get a denser grid of energies $E_i$ is not helpful since the solution is oscillating and non-positive at low energies 
(see Subsect.~\ref{sec:fits}). 

2. {\em Tikhonov regularization}. This case differs from TSVD in that the non-zero matrix elements of $D^{\dagger}$ are defined as 
$D^{\dagger}_{ii}=d_i/(d_i^2+\delta),i\le\min(n_G,n_d)$. The regularization parameter 
$\delta$ is again selected from the discrepancy principle; for the considered problem 
it turns out to be of the order of $\delta=0.007$ (see Fig.~\ref{fig:regfits}, right). 
Reducing the value of $\delta$ or increasing the quadrature rank $n_G$ gives rise to unphysical oscillations and negative values $\lambda_i$, while increasing $\delta$ suppresses the statistical errors $\delta^{\rm st}\lambda_i$ but also ``flattens'' the energy dependence.

  \begin{table}[h!]
\caption{Values of $\lambda_i=\lambda_{\rm pO}(E_i)$ and $\delta^{\rm st}\lambda_i$
(in units $10^{10} \text{s}^{-1}$)
for node energies $E_i<0.2$ eV,
calculated from (\ref{eq:linsys}) for $n_G=10$ using TSVD and Tikhonov regularization.}
 \label{tab:regfits}
 \begin{tabular}{l
 @{\hspace{3mm}}c
 @{\hspace{3mm}}c
 @{\hspace{3mm}}c
 @{\hspace{3mm}}c
 @{\hspace{3mm}}c
 @{\hspace{3mm}}c
  }
  $E_i$ (eV)
& 0.0064 & 0.0249 & 0.0540 & 0.0913 & 0.1335 & 0.1771
 \\
 \hline
 $\lambda_i(\delta^{\rm st}\lambda_i)$ (TSVD)
 & 2.30(0.61) & 4.32(1.59) & 20.52(2.29) & 11.50(1.62) & 3.59(0.54) & 0.88(0.14)\\
 $\lambda_i(\delta^{\rm st}\lambda_i)$ (Tikhonov)
 & 1.82(0.53) & 5.96(1.18) & 18.64(1.57) & 9.43(2.19) & 2.84(1.16) & 0.69(0.29)
\end{tabular}
\end{table}
The numerical values of the muon transfer rate $\lambda_i=\lambda_{\rm pO}(E_i)$ for node energies in the range $E_i<0.2$ eV, calculated from Eq.~(\ref{eq:linsys}) for $n_G=10$, are given in Table~\ref{tab:regfits}. The two regularization methods produce close results. A drawback of the approach is the rather scarce grid of energies $E_i, i\le n_G$, limited by the small number of data points $n_d$. The parametric fit approach, presented in the next subsection, attempts to circumvent this shortcoming.

\subsection{Constrained parametric fits}
\label{sec:fits}

\subsubsection{Constraints and selection criteria}

The ``constrained fit'' approach to the evaluation of $\lambda_{\rm pO}(E)$ will consist in searching for the best parametric fit to the experimental values $\Lambda_k$ with fitting functions 
$\Lambda(T;\{p\})$, obtained by convolution with $f_{\rm MB}(E;T)$ of  
TFs $\lambda(E;\{p\})$, which comply with the model-independent restrictions imposed by theory on the asymptotic behavior of $\lambda_{\rm pO}(E)$ at small and large values of $E$.
 The ``best fit'' will be selected according to the following criteria (Cr)
\begin{description}
\item[Cr1: Lowest value of $\chi^2$], {\rm where}
 \begin{eqnarray}
 &&\chi^2=\sum\limits_{i=k}^{n_d} (\Lambda(T_k;\{p\})-\Lambda_k)^2/\sigma_k^2=\min,\\
 &&\Lambda(T;\{p\})=\int\limits_0^{\infty} dE\,f_{\rm MB}(E;T)\,\lambda(E;\{p\}),
 \label{eq:basp}
 \end{eqnarray}

\item[Cr2: Stability of the fit] {\rm in the sense that no qualitative changes occur in case a subset of data points is excluded from the data set.}
\item[Cr3: Smallest width of the ``confidence band'' $\delta\lambda(E;\{p\})$,] 
{\rm defined in \cite{bates} as}
\begin{align}
&\delta\lambda(E;\{p\})=Q(\alpha,n_{\rm df}) \sqrt{\chi^2/n_{\rm df}}\, 
\sqrt{(\mathbf{J}^{\rm T}\mathbf{C}\mathbf{J})},
\label{eq:bates}\\ 
&\mathbf{J}=\{J_k\},k=1,...,n_p,\ J_k=\partial\lambda(E;\{p\})/\partial p_k,
\nonumber
\end{align}
where $\mathbf{C}$ is the covariance matrix, and $Q(\alpha,n_{\rm df})$ is the Student's t-distribution quantile for two-sided confidence level $\alpha$ and $n_{\rm df}$ degrees of freedom. This subsidiary semi-qualitative criterion will only be applied to fits with close 
values of $\chi^2$; it is based on the observation that, for such fits, the broader confidence intervals of the fit parameters may be a signal of significant correlation between them, which in turn may be due to inadequate choice of the trial functions.   
\end{description}
On the ground of general results of scattering theory about the asymptotical behavior 
of the rate of muon transfer $\lambda_{\rm pO}(E)$, we impose the following model-independent constraints (Co) on the trial functions $\lambda(E;\{p\})$,
used in fitting the experimental data:  

\noindent{\bf Co1: Non-negativity.} This constraint follows from the definition of the muon transfer rate $\lambda_{\rm pO}(E)$:
\begin{align}
\lambda(E;\{p\})\ge0\text{ for all }E\ge0.
\label{eq:co1}
\end{align}

\noindent{\bf Co2: Wigner threshold law.} According to Wigner's threshold law \cite{wigner},  in the limit of zero collision energy the rate of muon transfer $\lambda_{\rm pO}(E)$ is approximately constant for $E<E_W$, where
$E_W$ is the range of validity of the Wigner law. 
This can be physically understood as dominance of the $s$-wave at low energies.
In the absence of quantitative theoretical estimates of the specific value $E_W$ for the muon transfer process in Eq.~(\ref{eq:transfer}), we refer to Ref.~\cite{dalgarno}, which shows that the low-energy behavior of the rate of electron transfer between hydrogen isotopes, predicted by the Wigner law, becomes visible at collision energies $E$ below $10^{-5}-10^{-6}$ eV, or  $10^{-7}-10^{-8}$ in atomic units. In analogy, one may expect that the ``flat behavior'' of $\lambda_{pO}(E)$
is displayed in the energy range below $E_W\lesssim 10^{-7}\varepsilon_{\mu}\sim10^{-3}$ eV, where $\varepsilon_{\mu}=m_{\mu}\,c^2\alpha^2\sim5$ keV
denotes the ``$\mu$-atomic unit of energy'';
the  numerical results of Refs.~\cite{dupays1,cdlin,romanov22} point at even slightly higher values of $E_W$. We therefore impose the constraint
\begin{align}
\label{eq:co2}
\begin{split}
&\lambda_0=\lim\limits_{E\to0}\lambda(E;\{p\})>0, 
\\
&\left|\lambda'_0\right|=\left|d\lambda(E;\{p\})/d E\right|\ll\lambda_0/E_W 
\text{ for }E<E_W.
\end{split}
\end{align}

\noindent{\bf Co3: Large energy asymptotics.} We are not aware of any dedicated studies of the asymptotical behavior
 of the muon transfer rate to higher electric charge atomic nuclei. The general
 treatment of this class of atomic processes in Ref.~\cite{mensh},
 however, shows that for collision energies $E$ of the order of or higher than
 $\varepsilon_{\mu}$ 
 the transfer rate is a slowly decreasing function of $E$. This leads to
\begin{align}
d \lambda(E;\{p\})/dE\le0\text{ for }E>\varepsilon_{\mu}
\label{eq:co3}
\end{align}
In addition to these general constraints, we impose the following two constraints, specific for the considered problem:   

\noindent{\bf Co4: Limited number of adjustable parameters.} We shall focus on trial functions with $n_p<n_d/2=5$. Because of the small number of data points $n_d=10$, fits with larger number of parameters will have too few degrees of freedom that may lead to instabilities and numerical artifacts. 

\noindent{\bf Co5: Smoothness of the trial functions.} There are no evidences of threshold phenomena or processes that would give rise to discontinuities or singularities of 
$\lambda_{\rm pO}(E)$ in the considered range of collision energies. The theoretical calculations in \cite{dupays1,cdlin,romanov22} also predict a smooth energy dependence. We therefore restrict our search to the class of $C^{\infty}$ trial functions.

\subsubsection{Probing different classes of trial functions}

Constraints Co4 and Co5 eliminate a large variety of TFs that could possibly comply with constraints Co1--Co3. Before proceeding with the search of the ``best fit'', on a few examples we briefly review the basic features of these excluded TFs.

\begin{figure}[h!]
 \begin{center}
 \includegraphics[width=0.99\textwidth]{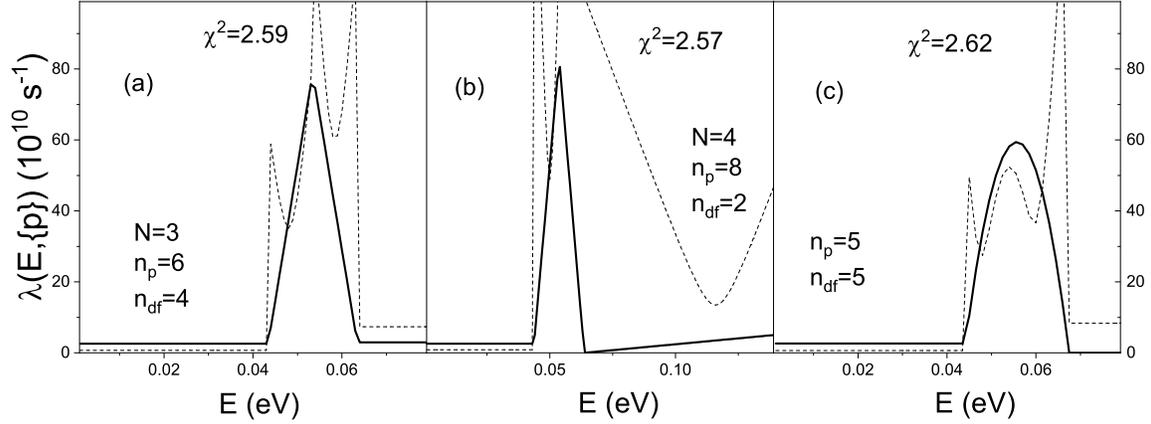}
 \caption{Fits of the experimental data in Table~\ref{tab:dataset} with piece-wise trial functions . The plots display the energy dependence of the trial functions $\lambda(E;\{p\})$ (solid line), together with the width of the confidence band $\delta\lambda(E;\{p\})$ (dashed), defined in Eq.~(\ref{eq:bates}), and {\bf shrunk by the factor of 5 to fit into the plot range. The real value of $\delta\lambda(E;\{p\})$ is 5 times larger than shown!}}
 \label{fig:linpiece}
 \end{center}
 \end{figure}

\begin{figure}[h!]
 \begin{center}
 \includegraphics[width=0.80\textwidth]{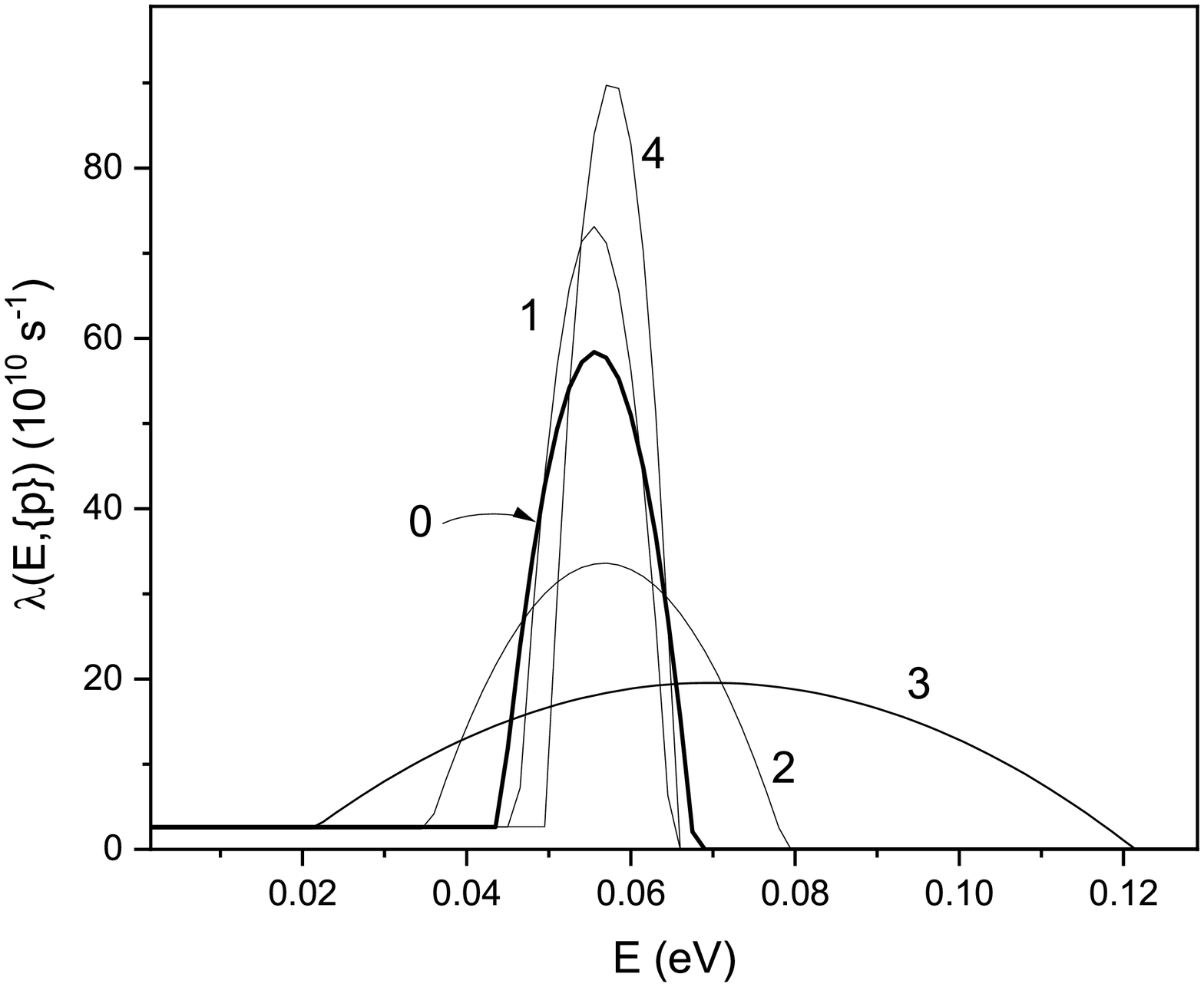}
 \caption{The trial function $\lambda(E;\{p\})$ of curve (c) in Fig.~\ref{fig:linpiece} with parameters $p_i, i=1,...,5$, computed by fitting the whole dataset in Table~\ref{tab:dataset} (curve "0") or data sub-sets obtained by excluding the data point in line $k, k=1,...4$ of the table (curves "$k$").}
 \label{fig:trpar}
 \end{center}
 \end{figure}

The simplest example are the piece-wise linear TFs involving $n_p=2N, N\ge2$ adjustable parameters $p_{k},k=1,\ldots,N$ (referred to as nodes) and 
$p_{k+N}=\lambda(p_{k};\{p\}),k=1,\ldots,N$ (function values at the nodes): 
\begin{align}
\lambda(E;\{p\})=\begin{cases}
p_{N+1},\text{ for }E<p_1 \\
\frac{(E-p_{k-1})}{p_{k}-p_{k-1}}p_{N+k}+
\frac{(p_{k}-E)}{p_{k}-p_{k-1}}p_{N+k-1}
\text{ for }p_{k-1}\le E<p_{k},\ k=2,\ldots,N\\
p_{2N},\text{ for }E\ge p_{N}.
\end{cases}
\end{align}
Non-negativity (Co1) is achieved by imposing the constraints 
$p_{k}\ge0,k=N+1,\ldots,2N$. 
The fits with $N=3$ and $N=4$ are shown on Fig.~\ref{fig:linpiece}, left and middle plots (a),(b). A possible generalization is the use of higher-order polynomial in some of the ``pieces'' (e.g. a parabola, as shown on plot (c).) The nonphysical discontinuities of the first derivative at the nodes $E=p_k, k=1,...,N$ lead to very broad confidence band around the node energies; this strongly suppresses the predictive potential of these fits. Most important of their shortcomings, however, is that, due do the small number of data $n_d$, the  number of degrees of freedom $n_{\rm df}$ is low; this leads, in turn, to instabilities, as illustrated on Fig.~\ref{fig:trpar}. 

In an attempt to overcome the problems related to first derivative discontinuities we have also probed cubic polynomial piece-wise (spline) trial functions of differentiability class  $C^2$, defined in a finite interval $0\le E\le E_{\rm max}$, and assumed to vanish outside of it. The number of parameters of such trial functions is related to the number of ``pieces'' $N$ as $n_p=2N+2$. For $N=2$ and $N=3$ the trial splines that minimize 
$\chi^2$ are incompatible with the non-negativity constraint Co1, while fits with higher $N$, for which the number of adjustable parameters $n_p$ approaches or exceeds the number of data points $n_d$, become unstable. On the basis of these considerations we conclude that piece-wise trial functions are inappropriate in fits of data sets with number of data $n_d$ as low as $n_d=10$. This may be considered as justification {\em a posteriori} of the adopted constraints
Co4 and Co5.

To comply with constraints Co1 and Co3 we selected trial functions that, for large values of $E$, asymptotically approach a non-negative constant:
$\lim_{E\to\infty}\lambda(E;\{p\})=\lambda_{\infty}\ge0$. 
We probed three kinds of trial functions: ``type 1'' for which, for large $E$, 
$\left|\lambda(E;\{p\})-\lambda_{\infty}\right|\sim {\rm const}. \exp(-\kappa E^2), 
\kappa>0$, ``type 2'' with  
$\left|\lambda(E;\{p\})-\lambda_{\infty}\right|\sim {\rm const}. \exp(-\kappa E), 
\kappa>0$, and ``type 3'' with 
$\left|\lambda(E;\{p\})-\lambda_{\infty}\right|\sim {\rm const}/E^{\alpha}, \alpha\ge0$. 

\begin{figure}[ht!]
 \begin{center}
 \includegraphics[width=0.99\textwidth]{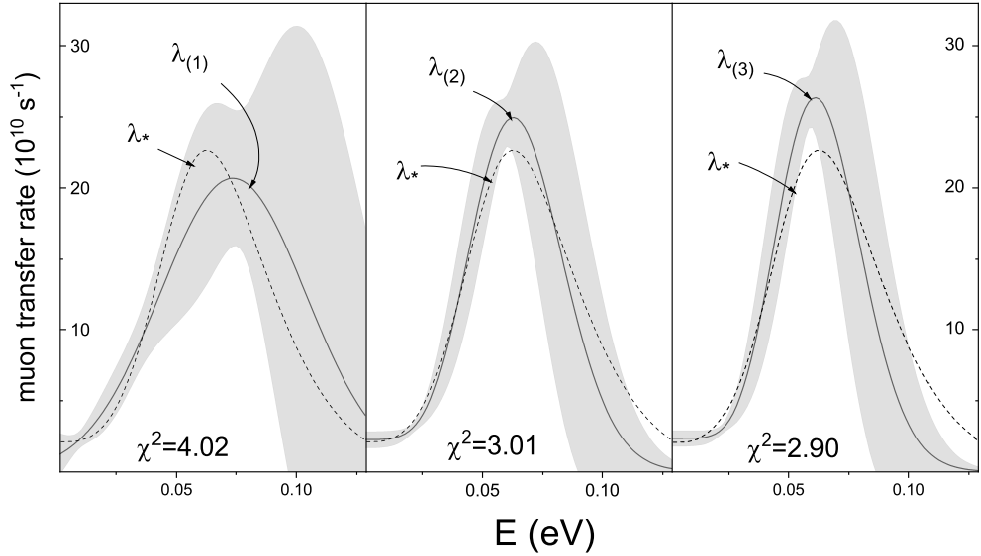}
 \caption{Energy dependence of the muon transfer rate $\lambda(E;\{p\})$ for the best fits of type 1. Left: 3-parameter Gaussian trial function of 
 Eq.~(\ref{eq:gauss.3pars}); middle: 4-parameter TF of Eq.~(\ref{eq:gaussex1}) with $\alpha_2=5$; right: same, with $\alpha_2=6$. The shadowed area represents the confidence band for 95\% CL. The dashed line is the best fit $\lambda_{*}(E;\{p\})$,
 defined in Eq.~(\ref{eq:bestfit*}).}
 \label{fig:type1}
 \end{center}
 \end{figure}

The family of trial functions of type 1 is initially taken in the form 
\begin{align}
&\lambda(E;\{p\})=p_3\left(1+\sum\limits_{k=2}^N (E/p_{k+2})^{\alpha_k}\right)
\exp(-(E-p_{1})^2/p_{2}^2)+p_{N+3},\ N=2,3,...
\label{eq:type1}
\end{align}
 where $\alpha_k\ge0$ are non-negative pre-selected fixed power exponents.
 This allows to evaluate the convolution with the Mawell-Boltzmann distrubution $f_{\rm MB}$ in Eq.~(\ref{eq:bas}) in closed form \cite{ryzhik,boya} that speeds up numerical optimization. 
The simplest 3-parameter TF of this type  
\begin{align} 
\lambda_{(1)}(E;p_1,p_2,p_3)=p_3\exp(-(E-p_1)^2/p_2^2)
\label{eq:gauss.3pars}
\end{align}
is strictly positive for any $E\ge0$ and leads to a reasonably good fit of the experimental data with $\chi^2=4.02$, which approximately satisfies the  Wigner's threshold law constraint Co2. 
To achieve better agreement with constraint Co2, we consider a modification of the 
family of the trial functions of Eq.~(\ref{eq:type1}):
\begin{align}
&\lambda(E;\{p\})=
p_3\left(1-\frac{2p_1 E}{p_2^2}+\frac{E^2}{p_2^2}
\left(1+\frac{2p_1^2}{p_2^2}\right)
+\sum\limits_{k=2}^N \frac{E^{\alpha_k}}{p_{k+2}^{\alpha_k}}\right)
\exp\left(-\frac{(E-p_1)^2}{p_2^2}\right)+p_{N+3}
\label{eq:gaussex1}
\end{align}
that includes additional terms, but no extra parameters. The extra terms guarantee that 
$\lambda'(0;\{p\})=\lambda''(0;\{p\})=0$ in agreement with Co2. 
Among the 4-parameter modified trial functions with $N=2$ the lowest values of $\chi^2$ are returned for $\alpha_2=5$ and 6; the corresponding TFs are denoted as 
$\lambda_{(2)}(E;\{p\})$ and 
$\lambda_{(3)}(E;\{p\})$, respectively. The above three TFs 
$\lambda_{(n)}(E;\{p\}), n=1,2,3$, shown on Fif.~\ref{fig:type1} will be retained in the short list of candidates for best fit of the FAMU experimental data. 
Extending the sum in Eq.~(\ref{eq:gaussex1}) to $N>2$ power terms or adding an intercept term $p_{N+3}$ returns fits with a bit lower $\chi^2$, but - similar to the ``truncated parabola fit'' on Fig.~\ref{fig:trpar} - unstable in the sense of Cr2. Such solutions once again justify the adoption of constraint Co4, and will not be considered in further analysis.

Type 2 trial functions are initially taken in the following form:
\begin{align}
&\lambda(E;\{p\})=p_1\left(1+\sum\limits_{k=2}^N (E/p_k)^{\alpha_k}\right)
\exp(-E/p_{N+1})+p_{N+2}.
\label{eq:type2g}
\end{align}
Similar to Eq.~(\ref{eq:gaussex1}), in order to comply 
with Wigner's threshold law we  modify them in a way to guarantee that
$\lambda'(0;\{p\})=\lambda''(0;\{p\})=0$ (assuming that $\alpha_k>2$):
\begin{align}
&\lambda(E;\{p\})=p_1\left( 1+E/p_{N+1}+E^2/(2p_{N+1}^2)+
\sum\limits_{k=2}^N (E/p_k)^{\alpha_k}\right)
\exp(-E/p_{N+1})+p_{N+2}.
\label{eq:type2m}
\end{align}
 \begin{figure}[t!h]
 \begin{center}
 \includegraphics[width=0.99\textwidth]{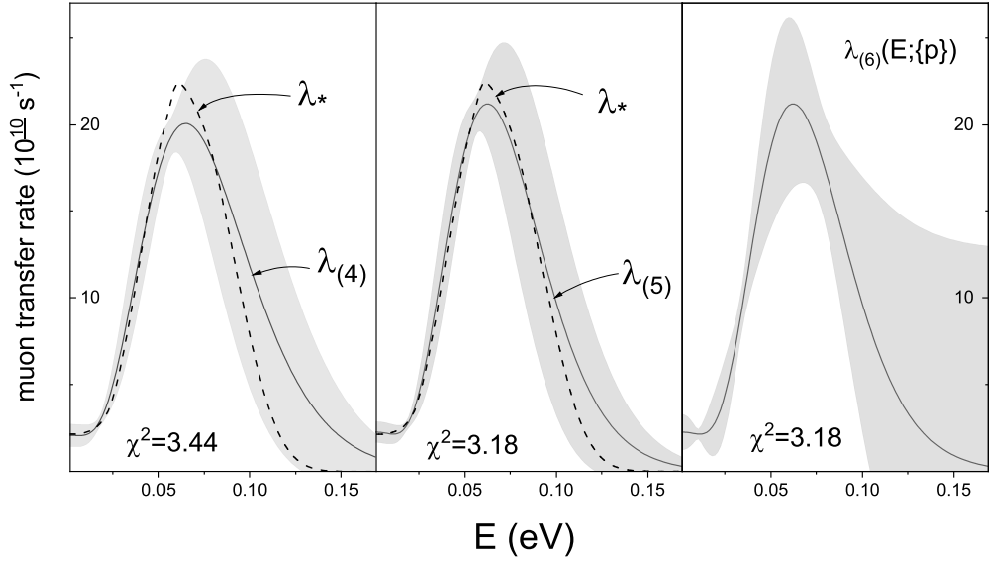}
 \caption{Energy dependence of the muon transfer rate $\lambda(E;\{p\})$ for the best fits of type 2. Left: 4-parameter TF with intercept, $\alpha_2=4$, see Eq.~(\ref{eq:gaussex1}); middle and right: 3-parameter TF without intercept, $\alpha_2=5$ and 6, respectively. The shadowed area represents the confidence band for 95\% CL. The dashed line is the best fit $\lambda_{*}(E;\{p\})$,
 defined in Eq.~(\ref{eq:bestfit*}).}
 \label{fig:type2}
 \end{center}
 \end{figure}
Out of the 3-parameter trial functions of type 2 we single out 
$\lambda_{(4)}(E;\{p\})$ and $\lambda_{(5)}(E;\{p\})$ with $N=2$ and $\alpha_2=5$ and  $\alpha_2=6$, respectively, as producing the fits with lowest $\chi^2$ 
(see Fig.~\ref{fig:type2}). 
We have also considered 4-parameter trial functions involving an intercept $p_4$.
Comparison of $\lambda_{(5)}((E;\{p\})$ and  $\lambda_{(6)}((E;\{p\})$, for both of which $N=2$ and $\alpha_2=6$, and which differ only by the presence of $p_4$ in the latter, shows that adding the intercept term does not lower the value of $\chi^2$ but significantly increases the width of the confidence band. 
This is due to the very large confidence interval of the intercept parameter $p_4$ 
and once again shows that the observable muon transfer rate $\Lambda(T)$ for 
$T\lesssim 300$ K  is uncorrelated with the energy dependence of the latter at epithermal or higher energies. 
Accordingly, trial functions with intercept, such as 
$\lambda_{(6)}((E;\{p\})$, will not be added to the short list. 
Note that $\lambda_{(6)}((E;\{p\})$ may also serve as an example of the applicability of criterion Cr3: out of two similar TFs $\lambda_{(5)}((E;\{p\})$ and  $\lambda_{(6)}((E;\{p\})$ with close $\chi^2$ we reject the fit with broader confidence band.
Trial functions involving $N>2$ terms in the sum in Eq.~(\ref{eq:type2m}) or higher powers $\alpha_k$ prove to either break constraints Co1/Co2 or return higher $\chi^2$ and will not be considered in further analysis either.
The parameters of the selected five best trial functions are given in Table~\ref{tab:short}. 

  \begin{table}[h!]
\caption{The values of $\chi^2$, of the predicted muon transfer rate at zero energy 
$\lambda_0$, and of the optimized parameters and their standard error for the 
selected trial functions in the ``short list'', defined in the previous section. The numerical values of the parameters are  in units meV=$10^{-3}$ eV, except for the boxed values, which are in units $10^{10}$ s$^{-1}$.}
 \label{tab:short}
 \begin{tabular}{c@{\hspace{2mm}}c@{\hspace{2mm}}c@{\hspace{2mm}}c@{\hspace{2mm}}c
 @{\hspace{8mm}}r@{\hspace{3mm}}r
 @{\hspace{8mm}}r@{\hspace{3mm}}r
 @{\hspace{8mm}}r@{\hspace{3mm}}r
 @{\hspace{8mm}}r@{\hspace{3mm}}r
  }
 label & $\chi^2$ & $n_p$ & $\lambda_0$ & $\alpha_2$ &
 $p_1$ & $\delta p_1$ &
 $p_2$ & $\delta p_2$ &
 $p_3$ & $\delta p_3$ &
 $p_4$ & $\delta p_4$ \\
     \hline
 $\lambda_{(1)}$ & 4.02 & 3 & \fbox{1.10} & 
& 73.7 & 19.6 & 43.0 & 14.1 & \fbox{20.7} & \fbox{2.9} \\
 $\lambda_{(2)}$ & 3.01 & 4 & \fbox{2.26} & 5 
& 9.96 & 4.8 & 36.9 & 7.1 & \fbox{2.44} & \fbox{0.51} & 26.4 & 4.2 \\
 $\lambda_{(3)}$ & 2.90 & 4 & \fbox{2.35} & 6 
& 3.03 & 7.8 & 34.8 & 6.6 & \fbox{2.37} & \fbox{0.52} & 25.8 & 3.8 \\
 $\lambda_{(4)}$  & 3.44 & 3 & \fbox{2.09} & 5 &
\fbox{2.09} & \fbox{0.64} & 15.4 & 3.3 & 13.1 & 2.4 \\
 $\lambda_{(5)}$ & 3.18 & 3 & \fbox{2.28} & 6 &
 \fbox{2.28} & \fbox{0.60} & 15.9 & 2.9 & 10.5 & 1.7 
\end{tabular}
\end{table}

Finally, as TFs of type 3, we probed a variety of Pad\'{e} approximants in the form 
\begin{align}
\lambda(E;\{p\})=p_0\frac{1+\sum\limits_{i=1}^m p_i E^{\alpha_i}}
{1+\sum\limits_{i=1}^n p_{i+m} E^{\beta_i}},
\label{eq:type3}
\end{align}
where $\alpha_i$ and $\beta_i$ are pre-selected positive integers, and $p_i,i=0,...,m+n$ are 
$m+n+1$ adjustable parameters. We did not find, however, any stable fit of type 3, involving up to $n_p=4$ parameters, which complies with constraint Co1 (non-negativity) and returns a competitive value of $\chi^2$. (Note that the ``best fit'' $\lambda_*^{\rm app}(E)$ defined in 
Eq.~(\ref{eq:finalfit}), which has the shape of a 6-parameter Pad\'{e} approximant, will be calculated as an approximation to the weighed average of the selected fits $\lambda_{\rm(1)}$ -- $\lambda_{\rm(5)}$, {\em not} to the experimental data.)

\section{Results and discussion}
 \label{sect3}

\subsection{Best fit and uncertainties.}
\label{sec:bestfit}

Out of the broad variety of trial functions, investigated in the ``constrained fit approach'' in Section \ref{sec:fits}, we have selected a ``short list'' $\cal{S}$ of 5 best fits (see Table~\ref{tab:short}) that comply with all constraints Co1--Co5, return the lowest values 
of $\chi^2$ in their class of TFs, and satisfy the selection criteria Cr2--Cr3.
 As long as the predicted energy curves for all 5 fits 
 diverge above 100 meV, we assume $E\le100$ meV as range of validity of our analysis. 
 Within this range, we assume as ``best fit'' 
 to the energy dependence of the muon transfer rate the weighed mean of the trial functions in the short list with weights, expressed in terms of the corresponding $\chi^2$ values:
\begin{align}
&\lambda_{*}(E;\{p\})=
\sum\limits_{k=1}^5 \omega_k(E)\lambda_{(k)}(E;\{p\}),\  \ 
\omega_k(E)=(\chi^2_{(k)})^{-1}
/\sum_{k'=1}^5 (\chi^2_{(k)})^{-1}.
\label{eq:bestfit*}
\end{align}    
 \begin{figure}[h!]
 \begin{center}
 \includegraphics[width=0.9\textwidth]{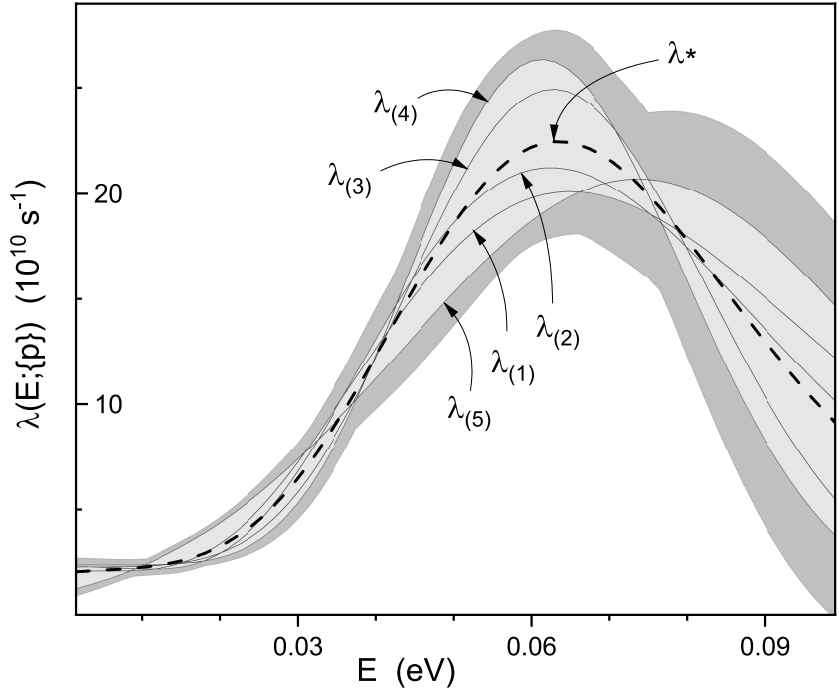}
 \caption{The five selected fits in the short list $\cal{S}$ (see Table~\ref{tab:short}), the best fit $\lambda_*(E;\{p\})$ of Eq.~(\ref{eq:bestfit*}) (thick dashed line), the model uncertainty band (light gray shadowed), 
 and the statistical uncertainty band (dark-gray-shadowed). }
 \label{fig:envelope5}
 \end{center}
 \end{figure}
Figure~\ref{fig:envelope5} illustrates the energy dependence of the rate of the muon transfer process (\ref{eq:transfer}) for the selected five fits. 
We interpret the envelope of the  curves of $\cal{S}$ (light-gray-shadowed on Fig.~\ref{fig:envelope5})  as ``model uncertainty band'' of the predicted best fit $\lambda_{*}(E;\{p\})$, and define the model uncertainty of $\lambda_{*}(E)$ as 
\begin{align}
\delta^{\rm M}\lambda_{*}(E;\{p\})=\big(
\max\limits_{i\in\cal{S}}\lambda_{(i)}(E;\{p\})\!-\!
\min\limits_{i\in\cal{S}}\lambda_{(i)}(E;\{p\})
\big)/2,
\nonumber
\end{align}
while the conservative estimate of the statistical uncertainty 
$\delta^{\rm st}\lambda_*(E;\{p\})$ 
(dark-gray-shadowed on Fig.~\ref{fig:envelope5}) is given by
\begin{align}
\delta^{\rm st}\lambda_*(E;\{p\})\le
\sum_{k=1}^5 \omega_k(E)\,\delta\lambda_{(k)}(R;\{p\}).
\label{eq:statband}
\end{align}
The model uncertainty of $\lambda_{*}(E;\{p\})$ does not exceed fractionally 30\% for $E<25$ meV, 20\% for $25<E<80$ meV, and 60\% for $80<E<100$ meV. The statistical uncertainty $\delta^{\rm st}\lambda_{*}(E;\{p\})$ is fractionally below 15\% for $E<80$ meV, and below 40\% for $E$ up to 100 meV. Conservatively, we define the total uncertainty as
  $\delta^{\rm tot}\lambda_{*}(E;\{p\})=
   \delta^{\rm M}\lambda_{*}(E;\{p\})+\delta^{\rm st}\lambda_{*}(E;\{p\})$. 
The total uncertainty is below 30\% for $E<80$ meV, and increases to 90\% for $E=100$ meV. 
We need to emphasize that the model uncertainty - unlike the statistical one - is not rigorously determined: the existence of smooth trial functions with asymptotical behavior compliant with constraints Co1--Co3 and leading to lower $\chi^2$, which lie outside the model uncertainty band on Fig.~\ref{fig:envelope5} cannot be ruled out. 
The  results presented here should be taken as semi-qualitative 
estimate of the systematic uncertainty of $\lambda_{*}(E;\{p\})$. 

 \begin{figure}
 \begin{center}
 \includegraphics[width=0.9\textwidth]{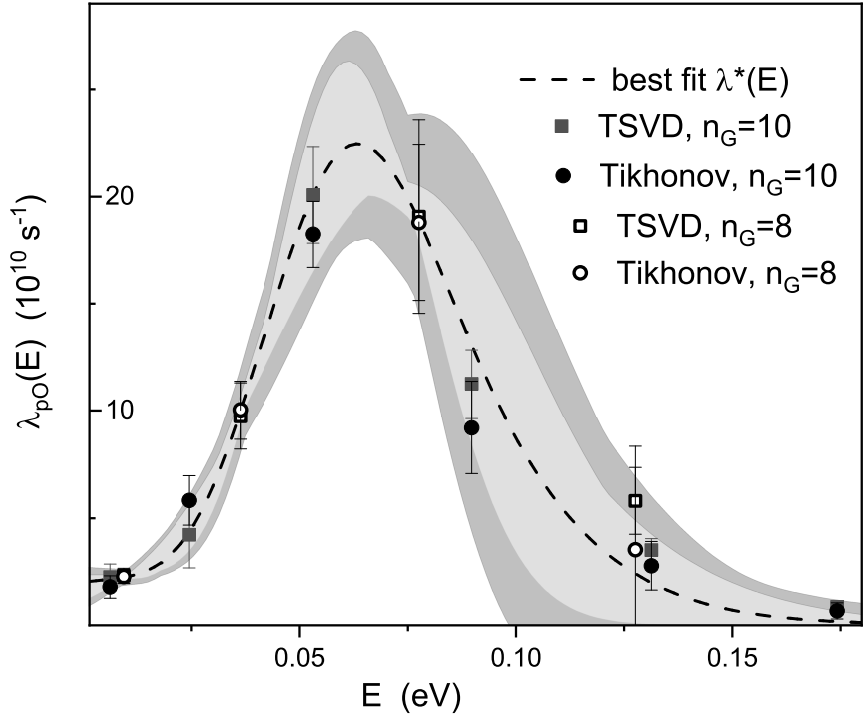}
 \caption{
 The best fit $\lambda*(E;\{p\})$ of Eq.~(\ref{eq:bestfit*}) (thick dashed line), the model uncertainty band (light gray shadowed), the statistical uncertainty band (dark gray shadowed), and the values of $\lambda_i = \lambda_{\rm pO}(E_i)$ and $\delta^{\rm st}\lambda_i$ for node energies $E_i<0.2$ eV, calculated from (\ref{eq:linsys}) using TSVD and Tikhonov regularization for $n_G=10$ (see Table~\ref{tab:regfits}) and $n_G=8$.}
 \label{fig:reg-vs-best}
 \end{center}
 \end{figure}

The comparison with the model-independent solutions of Section~\ref{sec:reg} confirms the credibility of the estimates obtained in the parametric fit approach. 
Fig.~\ref{fig:reg-vs-best} shows that the values $\lambda_i=\lambda_{\rm pO}(E_i)$
calculated by STVD and Tikhonov regularization of the discretized inverse problem (\ref{eq:bas}) for $n_G=8$ and $n_G=10$ and $E_{\rm max}=0.3$ eV fit well into the model uncertainty band of the ``best fit''; the good agreement is true for any $6\le n_G\le10$ and 
$0.25<E_{\rm max}<0.40$ eV. 
In any of the approaches, the same sharp peak of 
$\lambda_{*}(E;\{p\})$ around 6 meV is displayed; this peak is well-distinguishable also on the plots of fits that were rejected for breaking Co5 (see Fig.~\ref{fig:linpiece}). Note also that the best fit 
$\lambda_{*}(E;\{p\})$ fits into the confidence band of any individual TF of the short list $\lambda_{(k)}(E;\{p\}),k=1,...,5$ (see Figs.~\ref{fig:type1},\ref{fig:type2}).

 In computations, in the energy range of interest $0<E<100$ meV the values of 
 $\lambda_*(E;\{p\})$ can be approximated with mean fractional error of the order of 1.1\% with
\begin{align} 
& \lambda_{*}^{\rm app}(E;\{p\})=
2.109\frac
{1-(E/47.79)^2+(E/22.06)^3}
{1+((E+4.690)/66.84)^6+(E/86.58)^{12}}
\label{eq:finalfit}
\end{align}
where $E$ is taken in units meV, and the rates are evaluated in units $10^{10}$ s$^{-1}$.

\subsection{Comparison with theory.} 

When proceeding to comparison with theory, we should keep in mind that all known calculations consider the muon transfer to an oxygen nucleus in scattering of $p\mu$ by an oxygen atom, not molecule, i.e. they return $\lambda^{\rm A}_{\rm pO}(E)$, which -- as explained in Sect.~\ref{sect2A} -- may be different from the rate $\lambda_{\rm pO}(E)$, determined in the present work. 

\begin{figure}
 \begin{center}
 \includegraphics[width=0.95\textwidth]{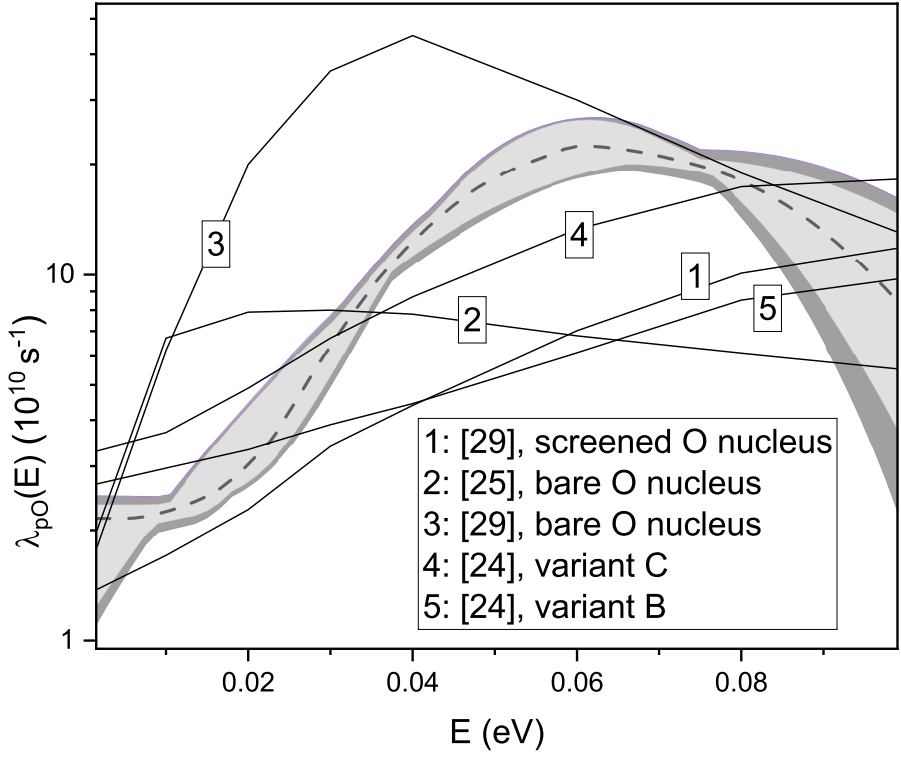}
 \caption{Comparison  of the results of the advanced computations of the rate of muon transfer to oxigen in Refs.~\cite{romanov22,dupays1,cdlin} with the experimentally determined energy dependence $\lambda_*(E;\{p\})$ (dashed) and its total uncertainty band composed of  model (light-gray-shadowed) and statistical uncertainties (dark-gray), for $E<100$ meV.}
 \label{fig:figure3}
 \end{center}
 \end{figure}

The theoretical results of Refs.~\cite{dupays1,cdlin,romanov22}
use various physical approximations that may be responsible for the observed quantitative differences between them: models of the electron structure of the oxygen atom in \cite{cdlin,romanov22}, or neglect of the latter in \cite{dupays1}, neglect of the spin interactions and of the O$_2$ molecular structure, etc.
Fig.~\ref{fig:figure3} juxtaposes the energy dependence derived in the present work with the theoretical curves for the muon transfer to a ``bare oxygen nucleus'' \cite{dupays1,cdlin} and a screened one \cite{cdlin,romanov22}; the latter were digitized from Fig.~2 of Ref.~\cite{dupays1}, Fig.~6 of Ref.~\cite{cdlin}, and 
Fig.~3 of Ref.~\cite{romanov22}.
All curves have a pronounced peak in the investigated energy range, which
might be related to a $p$-wave resonance. 
The peak of our best fit $\lambda_{*}(E;\{p\})$ is positioned  between the peak predicted by ``bare O-nucleus'' calculations, and the ``screened nucleus'' peak. 
The values of the computed muon transfer rate are in general outside the total 
uncertainty band of the experimental curve, and only approach it at lower 
energies and above 80 meV, where the experimental uncertainty increases. As a whole, the ``distance'' between the results of the various theoretical calculations 
and $\lambda_{*}(E;\{p\})$ or 
among themselves significantly exceeds the experimental uncertainty. Closest to $\lambda_*(E;\{p\})$ are the results of the recent work \cite{romanov22}, version C; for the thermal energies at 300 K, $E\sim39$ meV, they are in good agreement.
In the zero-energy limit, most of the calculations converge to close values within the uncertainty band around the experimental curve, with the exception of \cite{romanov22}, version C, which predicts a higher value. 

\subsection{Comparison with experiment.} 

\begin{figure}[h]
 \begin{center}
 \includegraphics[width=0.90\textwidth]{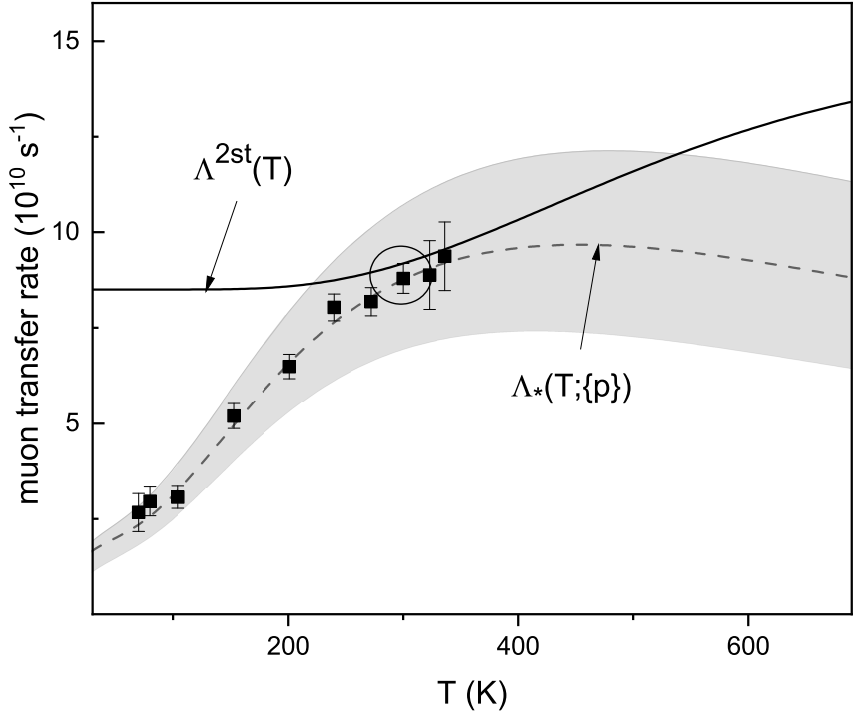}
 \caption{Comparison of the temperature dependence $\Lambda_{\rm pO}^{\rm 2st}(T)$ 
 of the muon transfer rate to oxygen predicted by the two-step model of 
 Ref.~\cite{werth0} with the best fit to the FAMU experimental data 
 $\Lambda_*(T;\{p\})$. The shadowed area represents the total (model $+$ statistical) 
 uncertainty band of $\Lambda_*(T;\{p\})$. The circle denotes the area where the 
 two curves are expected to overlap.}
 \label{fig:exper-new}
 \end{center}
 \end{figure}

The FAMU collaboration is the first to directly investigate the energy dependence of the rate $\lambda_{\rm pO}(E)$ of muon transfer from hydrogen to oxygen;
the preceding studies in Ref.~\cite{werth0} were aimed at distinguishing the contribution from thermal and epithermal muonic atoms.
Strictly speaking, the ``two-step'' function $\lambda_{\rm pO}^{\rm 2st}(E^{\rm L})$, reported in  Ref.~\cite{werth0}, is not describing the dependence of the muon transfer rate to oxygen on the center-of-mass collision energy $E$, but the dependence on the lab-frame energy $E^{\rm L}$ of the $p\mu$ atom at temperature 300 K, obtained by averaging  $\lambda_{\rm pO}^{\rm 2st}(E)$
 over the lab frame thermal  kinetic energy $E_{\rm O}^{\rm L}$
 of the oxygen molecule:
  \begin{equation}
  \lambda^{\rm 2st}_{\rm pO}(E^{\rm L},T)\!=\!
  \frac{2}{\sqrt{\pi\varepsilon_T^3}}\int_0^{\infty}
  \!dE_{\rm O}^{\rm L}\sqrt{E_{\rm O}^{\rm L}}
  e^{-E_{\rm O}^{\rm L}/\varepsilon_T}\,
  \lambda_{\rm pO}^{\rm 2st}(E),\ \ \varepsilon_T=k_BT.
  \label{eq:llab}
  \end{equation}
 Because of the large oxygen molecule mass compared to the mass of $p\mu$, however, 
 $E^{\rm L}\approx E$ and $\lambda^{\rm 2st}_{\rm pO}(E^{\rm L},T)\approx
 \lambda_{\rm pO}^{\rm 2st}(E)$. 
 On Fig.~\ref{fig:exper-new} we juxtapose the temperature dependence 
 $\Lambda_{\rm pO}^{\rm 2st}(T)$ 
 of the muon transfer rate to oxygen, generated from $\lambda_{\rm pO}^{\rm 2st}(E)$
 by convolution with the Maxwell-Boltzmann distribution (see Eq.~(\ref{eq:bas}))
 with the best fit to the FAMU experimental data 
 $\Lambda_*(T;\{p\})$, generated in the same way from $\lambda_*(E;\{p\})$. 
 The two curves are very close at 300 K, around the only then-available experimental value. Below 300 K the two-step model predicts a flat behavior in contrast with the observed decrease of the muon transfer rate with temperature. Above 300 K the two-step model
predicts an increase of the rate up to about 1000 K - a temperature range that is 
inaccessible with the FAMU experimental method and stands outside the range of validity of the fit developed in the present work.

\section{Summary and conclusions}
 \label{sect4}

The accurate determination of the energy dependence $\lambda_{\rm pO}(E)$ of the rate of muon transfer from muonic hydrogen to oxygen in the thermal and near-epithermal range has a two-fold motivation:
to dissolve the long-standing and persisting ambiguity around the sharp raise of $\lambda_{\rm pO}(E)$ and to establish reliable references for the methods of quantitative description of charge-exchange processes involving ordinary and exotic atoms, and
to provide firm ground for the optimization of the FAMU experiment
and in planning of further experiments using this technique.

 {\em Verification of the FAMU experimental method of measuring the hyperfine splitting in the ground state of muonic hydrogen.}  Muonic hydrogen $p\mu$ is one of the few exotic hydrogen-like atoms,
 whose lifetime is long enough to allow for precision spectroscopy.
 The ground-state hyperfine splitting of $p\mu$, 
 $\Delta E^{\rm hfs}\sim 0.182$ eV,
 turns out to be in the infra-red optical range, thus enabling the application of laser spectroscopy techniques.
 A number of experimental proposals for the measurement of $\Delta E^{\rm hfs}$
 have been put forward in recent years
 \cite{MMM,nimb12,jinst16,jinst18,epja,crema-las,crema-new,japs-las}. This was stimulated by the
 need of new data on the proton electromagnetic structure that had become an issue with the proton charge radius determination from the Lamb shift in muonic hydrogen \cite{pohl}.
In all these proposals the muonic hydrogen atom is being excited from the ground singlet to the triplet state with a laser, tunable around the resonance frequency 
$\Delta E^{\rm hfs}/h\sim44$ THz; the experimental methods differ by the signature used to detect the laser-induced transitions.
In the FAMU experimental method $p\mu$ propagates in a gaseous mixture of hydrogen and oxygen.
Collisions of $p\mu$ with oxygen lead to the reaction (\ref{eq:transfer}); the events of muon transfer are signaled by the characteristic X-rays emitted during the de-excitation of the muonic oxygen. The observable in the FAMU experiment is the time distribution of these events.
The $p\mu$ atoms that have been excited to the triplet state with a laser pulse are accelerated after the de-excitation in subsequent collisions with the surrounding 
H$_2$ molecules by nearly 0.12 eV; the atoms carry the released energy away as kinetic energy. Since the rate of muon transfer varies with the $p\mu$ kinetic energy $E$, the observed time distribution of the characteristic X-rays is perturbed as compared to the time distribution in absence of laser radiation; the resonance frequency is  recognized by the maximal response of the X-ray time distribution. (For details see Refs.~\cite{nimb12,jinst16,jinst18,epja}).
The efficiency of this method 
of detecting the events of laser-induced hyperfine excitation of $p\mu$
depends on how much the rate of muon transfer from accelerated 
$p\mu$ atoms exceeds the transfer rate from thermalized atoms. 
The hydrogen-oxygen mixture had been selected for the FAMU method because of the evidences in \cite{werth0,werth} for a sharp energy dependence of 
$\lambda_{\rm pO}(E)$ at thermal and near epithermal energies that is not observed in other gases. 
The results of the present work establish a raise by nearly an order of magnitude of the muon transfer rate to oxygen with energy from $E\sim10$ to $\sim\,70$ meV, far above the threshold considered in earlier simulations \cite{MMM}.  
Moreover, the knowledge of the detailed energy dependence of $\lambda_{\rm pO}(E)$ provides 
the information needed -- together with the scattering cross sections of muonic hydrogen elastic scattering \cite{atlas,atlas-n} -- for reliable modeling of the experiment and fine-tuning the experimental conditions for maximal efficiency - a task that is, however, out of the scope of this paper.

{\em Reference dataset for computations of charge exchange processes}.
In the energy range $0.01<E<0.08$ eV the total (model and statistical) fractional 
 uncertainties are below 30\% and the values of $\lambda_{*}(E;\{p\})$ are  reliably related to the experimental data.
 These results offer the rare opportunity to calibrate the computational quantum mechanical methods for the quantitative description of low energy inelastic scattering of light atomic systems.

 It should be emphasized that the experimental method for the determination of the energy dependence of the rate of muon transfer by repeated measurements in thermal equilibrium at different temperatures is directly applicable to the study of muon transfer to other gases. In the absence of specific restrictions on the range of investigated temperatures as in the case of hydrogen-oxygen gas mixture, the range of validity of the experimentally determined energy dependence of the muon transfer rate could also be extended compared to the oxygen case considered here. Since the constraints on the trial functions stem from most general principles, the class of trial functions used in the present work is expected to be appropriate in these studies as well.

 \begin{acknowledgments}
 M.S., D.B. and P.D. acknowledge the support of Bulgarian
 National Science Fund Grant No. KP-06-N58/5. D.B is grateful to K. Boyadzhiev for helpful discussions. 
 \end{acknowledgments}

 \end{document}